\documentclass[acmlarge]{acmart}
\makeatletter
\newcommand{\myconfshort}{\acmConference@shortname}
\newcommand{\myconffull}{\acmConference@name}
\newcommand{\myconfdate}{\acmConference@date}
\newcommand{\myconfloc}{\acmConference@venue}
\AtBeginDocument{
  \fancypagestyle{firstpagestyle}{
    \fancyhead{}%
    \fancyfoot[C]{}%
  }
  \fancyhf{}
  \fancyhead[LO]{\@headfootfont\shorttitle}%
  \fancyhead[RE]{\@headfootfont\@shortauthors}%
  \fancyhead[LE]{\@headfootfont\footnotesize \myconfshort, \myconfdate, \myconfloc}%
  \fancyhead[RO]{\@headfootfont\footnotesize \myconfshort, \myconfdate, \myconfloc}%
  \fancyfoot[C]{}%
}
\makeatother
\acmBooktitle{\conffull\@ (\confshort), \confdate, \confloc}

\copyrightyear{2026}
\acmYear{2026}
\setcopyright{cc}
\setcctype{by}
\acmConference[FAccT '26]{The 2026 ACM Conference on Fairness, Accountability, and Transparency}{June 25--28, 2026}{Montreal, QC, Canada}
\acmBooktitle{The 2026 ACM Conference on Fairness, Accountability, and Transparency (FAccT '26), June 25--28, 2026, Montreal, QC, Canada}
\acmDOI{10.1145/3805689.3812335}
\acmISBN{979-8-4007-2596-8/2026/06}

\AtBeginDocument{%
  }

\hypersetup{colorlinks=true, linkcolor=blue, urlcolor=blue, citecolor=blue}

\begin{document}

\title{A Neuro-Symbolic Framework for Legal Accountability in Public-Sector AI}

\author{Allen Daniel Sunny}
\affiliation{
  \institution{University of Maryland}
  \city{College Park}
  \state{MD}
  \country{USA}
}
\email{allens@umd.edu}

\author{Ido Sivan-Sevilla}
\affiliation{
  \institution{The Hebrew University of Jerusalem \& University of Maryland}
  \city{Jerusalem}
  \country{Israel}
}
\email{sevilla@cs.huji.ac.il}



\begin{abstract}
  Automated decision-making systems are increasingly used by public agencies to govern access to welfare. Accountability in these settings is not achieved through access to model internals, but by the explanations provided to applicants i.e. documents that function as legal justifications and sites of contestation. Little work has been done to examine whether such explanations are legally valid. This paper examines explanation-level accountability in public-sector AI systems by designing and implementing a neuro-symbolic framework to assess automated welfare eligibility determinations. We deploy a Large Language Model (LLM) to encode statutory eligibility rules into a formal ontology and use satisfiability-based verification to assess whether benefits explanations comply with the governing law. Applying this approach to California's CalFresh program, we analyze explanations and observe if our system detects legal mismatches. Our findings show that formal verification reveals violations of statutory requirements even when explanations appear reasonable or complete to human readers. We argue that explanation-level verification offers a distinct and necessary complement to existing approaches of algorithmic auditing and accountability, shifting attention from model behavior to the legal integrity of justificatory artifacts. We call for a shift from interpretability to auditability of public algorithms, as government explanations must do more than reveal statistical associations; they should articulate the legal basis for a decision so that affected individuals, oversight bodies, and administrative reviewers can assess whether the justification satisfies statutory criteria and procedural norms.
\end{abstract}

\begin{CCSXML}
<ccs2012>
   <concept>
       <concept_id>10010147.10010178.10010187.10010195</concept_id>
       <concept_desc>Computing methodologies~Ontology engineering</concept_desc>
       <concept_significance>500</concept_significance>
       </concept>
   <concept>
       <concept_id>10003456.10003462.10003588.10003589</concept_id>
       <concept_desc>Social and professional topics~Governmental regulations</concept_desc>
       <concept_significance>300</concept_significance>
       </concept>
   <concept>
       <concept_id>10010405.10010455.10010458</concept_id>
       <concept_desc>Applied computing~Law</concept_desc>
       <concept_significance>100</concept_significance>
       </concept>
 </ccs2012>
\end{CCSXML}

\ccsdesc[500]{Computing methodologies~Ontology engineering}
\ccsdesc[300]{Social and professional topics~Governmental regulations}
\ccsdesc[100]{Applied computing~Law}
\keywords{Algorithmic accountability, Explainable AI, Public-sector AI, Neuro-symbolic systems}

\maketitle

\section{Introduction}
Government agencies have increasingly turned to automation to maintain service delivery~\cite{calo_automated_2021}. The rapid advancement of artificial intelligence and machine learning has accelerated this shift, encouraging the adoption of automated decision systems for welfare benefits with the promise of efficiency and consistency~\cite{yeung_new_2023}. At the same time, integrating machine learning into public-benefit decision making introduces serious risks~\cite{bednarz_automated_2023}. Advanced predictive systems are often opaque, making it difficult or even impossible to understand how eligibility decisions are made. Decisions that cannot be explained cannot be meaningfully challenged, restricting applicants’ rights to contest errors that may deny them essential support. 

Automated eligibility systems,  whether predictive, rule-based, or hybrid, produce determinations whose internal logic is often inaccessible to the individuals they affect \cite{bovens_street-level_2002}. Eligibility criteria are expressed in complex legal language, and it is unrealistic to expect applicants to hold the knowledge needed to assess their eligibility explanations \cite{Escher_and_Banovic}. At the same time, the reasoning behind automated determinations is rarely surfaced in the explanations provided to applicants \cite{veale2019administration}. This disconnect means that explanations provided to applicants may appear plausible yet fail to reference the required legal authority or may misstate the conditions under which benefits should be granted. Notices of action, benefit determination letters, and similar administrative texts serve as the interface through which state authority is exercised and made contestable. These explanations must document how statutory criteria were applied, signal the legal basis for eligibility or denial, and structure the procedural rights of individuals who seek review or appeal \cite{citron_technological_2008}.

This governance framing stands in contrast to much of the explainable AI literature, which treats explanations as tools for interpretability, debugging, or user trust \cite{rudin_stop_2019}. XAI methods typically optimize for fidelity to model behavior or user comprehensibility \cite{doshi-velez_towards_2017}. Legal and institutional review evaluates explanations according to a different criterion: traceability to authoritative sources of law, coherence with procedural rules, and suitability for contestation in adversarial or quasi-judicial settings \cite{citron_technological_2008}. Interpretability is a necessary condition for accountability, but it is not sufficient in legal contexts, and bridging this gap requires moving from explaining models to auditing justifications. Such a shift requires formal representations of legal rules, mechanisms that map explanation content to those rules, and procedures that determine whether stated justifications satisfy applicable statutory requirements \cite{nay_law_2023}. Without a shared legal structure, neither applicants nor administrators can determine whether an automated decision is justified under the law. This gap highlights a fundamental accountability problem: explanations must be legally valid and not just interpretable, if automation is to preserve due-process rights in public-benefit administration \cite{eubanks_automating_2018}.

Previously, a similar formalization of welfare distribution was conducted by Escher and Banovic (2020) \cite{Escher_and_Banovic}. Their method, however, focused on the validity of benefits screening tools rather than the explanations provided to applicants. Our work addresses a distinct governance function: rather than producing eligibility determinations, we evaluate whether justifications already issued are legally adequate, supporting contestation and appeal rights that existing technical infrastructure rarely enables.

To address this gap, we build a neuro-symbolic framework to evaluate the legal accountability of public benefits programs and demonstrate its validity on California's SNAP program (CalFresh). We rely on literature about formal knowledge representation, ontology extraction, and neuro-symbolic frameworks (Section 2), describe our methodology and architecture (Section 3), and report model accuracy against the CalFresh case study (Section 4). We discuss our results (Section 5) and conclude (Section 6) with a call for a shift towards explanations-based public algorithmic accountability.

\section{Literature Review: Legal Reasoning for Welfare Distributions via Neuro-Symbolic Frameworks}

To automatically verify the legal basis for public benefit distributions, the paper brings together literature about knowledge representation (of the law) and ontology extraction and construction via formal methods. The paper uses Large Language Models (LLMs) to represent the law and builds the relevant ontology at scale, to then put everything together through a neuro-symbolic framework for legal verification of welfare distribution.

Knowledge representation is applied through an ontology, defined as a formal specification of how entities within a domain are categorized and related to one another \cite{gruber_translation_1993}. It articulates a shared conceptualization of reality, defining the classes (types of things), relations (how those things connect), and axioms (constraints that must hold true) that together describe a domain’s structure. Gruber defines an ontology as “an explicit specification of a conceptualization” \cite{gruber_translation_1993}, positioning ontologies as a bridge between human semantic categories and the formal structures required for computational reasoning in legally regulated domains. Over time, several representational standards have emerged to encode such knowledge structures, ranging from the Resource Description Framework (RDF) \cite{w3c1999rdfsyntax}, which expresses knowledge as subject–predicate–object triples, to the Web Ontology Language (OWL) \cite{w3c2012owl2}, which builds on RDF to support richer logical semantics based on Description Logic. 

From these standards, the field of representation of legal knowledge has emerged, seeking to separate the structure of legal concepts from the logic of their application. The emergence of legal knowledge representation happened through the development of ontologies which are formal, machine-readable vocabularies of legal entities and relations \cite{computational_law,British_Act_As_Logic}. Frameworks such as LKIF-Core (Legal Knowledge Interchange Format) \cite{hoekstra2007lkif}, LegalRuleML \cite{LegalRuleML_Design}, and other OWL-based models introduced  dependencies between rules. These ontological frameworks allowed legal knowledge to be represented with greater precision and interoperability, allowing reasoning engines to perform tasks such as compliance checking, conflict detection, and inference of legal consequences. In doing so, they transformed legal texts from static documents into structured knowledge bases that could be computationally queried and analyzed. This formalization provided a foundation for a new class of reasoning systems: once legal rules were expressed as explicit logical statements, they could be subject to automated verification and consistency checking. A related institutional effort is the Rules as Code movement, which advocates encoding legislation as machine-executable specifications at the drafting stage to improve administrative clarity and automation \cite{mohun_cracking_2020}.

The construction of ontological vocabulary against the way rules are expressed is a major challenge. Legal ontologies have traditionally been built through manual expert analysis — a process that is labor-intensive, difficult to maintain as statutes evolve, and prone to inconsistency when multiple analysts encode the same domain. Recent work has explored automating this process by extracting ontological structure directly from legal text. Concept extraction pipelines segment statutory provisions into minimal eligibility-relevant clauses and identify legally operative terms through noun-phrase extraction, named entity recognition, or embedding-based similarity matching \cite{gibaut_neurosymbolic_2023, wan_towards_2024}. Extracted concepts can then be organized into hierarchical class structures that reflect the normative organization of the source legislation, with de-duplication enforced through semantic similarity thresholds to prevent redundant or overlapping representations.

A key challenge in this process is maintaining alignment between extracted concepts and their authoritative legal sources. Without explicit traceability — linking each ontology term to a specific statutory citation — the resulting vocabulary risks becoming disconnected from the legislation it is meant to represent. This concern is particularly acute in welfare administration, where eligibility depends on precise statutory conditions and where misalignment between the ontology and the governing law could produce verification errors that harm affected individuals. Embedding-based methods offer a partial solution by enabling both de-duplication and structural validation. By projecting extracted concepts into a shared semantic space, it becomes possible to assess whether legally distinct eligibility domains — income, residency, citizenship — remain separable in the representation, a necessary condition for accurate downstream rule retrieval and verification.

Importantly, representing legal knowledge in formal structures is a necessary but insufficient condition for accountability. Reasoning systems operationalize these representations by evaluating whether specific facts, claims, or decisions satisfy encoded legal constraints. Satisfiability Modulo Theories (SMT) \cite{hutchison_satisfiability_2009} provide the foundational computational mechanism for this task. SMT solvers extend propositional logic with domain-specific theories — including arithmetic, temporal relations, and string comparison — enabling evaluation of both qualitative and quantitative legal conditions. For example, determining whether a household's income satisfies statutory thresholds while residency verification obligations are jointly met, requires reasoning across multiple constraint types simultaneously.

Systems such as Regorous \cite{governatori2015regorous} integrate legal ontologies with SMT-based constraint solving to perform process-level compliance auditing, while other approaches employ logical reasoners to infer legal consequences or detect rule conflicts \cite{judson_put_2024, khang_automating_defeasible_2022}. Judson et al. \cite{judson_put_2024} demonstrate how SMT-based oracles can investigate the reasoning behind automated decisions, combining formal verification with decision traceability. Collectively, this body of work shows that symbolic reasoning can render the structure of law computationally tractable for compliance tasks. Still, a common limitation persists: existing reasoning systems rely on exhaustive manual rule enumeration, require expert maintenance, and remain disconnected from unstructured natural-language texts that constitute the operative interface between agencies and affected individuals \cite{libal_legal_2023, khang_automating_defeasible_2022}. 

Large Language Models (LLMs) have introduced new possibilities for bridging natural-language legal texts and formal representations. Unlike earlier rule-based extraction methods that required hand-crafted templates or expert annotation, LLMs can process statutory language at scale and generate candidate logical formalizations from unstructured text. Recent work has explored 'LLM-Plus-Solver' architectures in which neural models extract candidate rules or explanations from natural language, and symbolic solvers check the satisfiability of those claims \cite{sadowski_explainable_2025}. These pipelines have been applied to visual reasoning \cite{dong_neural_2019}, scientific discovery \cite{trinh_alphageometry_2024}, and automated theorem proving \cite{rao2025neuraltheoremprovinggenerating}, demonstrating that neural interpretations can be systematically tested against external logical constraints rather than accepted at face value. In legal domains specifically, LLMs have been used to identify legally operative concepts, clause boundaries, and exceptions from regulatory texts \cite{noguer_i_alonso_automating_2024}, while related approaches combine rule extraction with constraint validation — for example, LLM + Z3 or LLM + Prolog pipelines — verifying whether the logic behind a textual explanation is complete and consistent \cite{calanzone_logically_2024, wang_python_2024}.

The reliability of LLM-based formalization, however, is critically dependent on the prompting strategy. Unconstrained generation frequently produces narratively plausible but syntactically invalid or semantically drifted outputs. Directed symbolic prompting — in which the model is explicitly instructed to produce solver-compatible logic using a controlled vocabulary — has been shown to dramatically improve formalization success rates, while vanilla and undirected approaches yield unreliable results even from frontier models \cite{calanzone_logically_2024, sadowski_explainable_2025}. These findings position LLMs not as autonomous legal reasoners but as extraction and translation components within larger verification pipelines — capable of scaling the formalization of statutory text, provided their outputs are validated against independent symbolic constraints.

The different components above can come together via Neuro-symbolic frameworks. Neuro-symbolic AI integrates neural learning with symbolic reasoning, combining the flexibility of data-driven methods with the formal guarantees of logic-based systems \cite {gibaut_neurosymbolic_2023}. In settings where behavioral errors carry significant consequences and correctness must be formally established, this integration enables outputs to be verified rather than merely interpreted.

Foundational architectures such as Logic Tensor Networks \cite{badreddine_logic_2022} and DeepProbLog \cite{manhaeve_deepproblog_2018} incorporate logical constraints directly into neural inference, ensuring that predictions respect domain-specific structure rather than relying on unconstrained statistical associations. More recent approaches connect LLMs to external verification systems, creating pipelines in which neural components generate structured representations and symbolic components evaluate them  \cite{calanzone_logically_2024, sadowski_explainable_2025}. In these architectures, the division of labor is deliberate: neural methods handle the variability and scale of natural-language inputs, while symbolic methods provide the deterministic reasoning required for verification and traceability.

In legal domains, this paradigm aligns directly with the requirements of procedural accountability. Neural systems can extract legally operative concepts and translate statutory clauses into formal constraints, while symbolic solvers verify whether these constraints are jointly satisfiable under given case facts \cite{wang_python_2024}. The resulting systems can detect not only that an inconsistency exists but also which specific legal provisions are implicated — transforming verification outputs into governance-relevant accountability signals.

Despite these advances, no existing neuro-symbolic system has been designed to verify the legal adequacy of the explanations that public agencies issue to affected individuals. Current approaches focus on decision-level compliance or rule-level consistency, but do not address whether the justificatory artifacts — notices of action, eligibility explanations, benefit determination letters — satisfy the statutory requirements that authorize them. This gap between decision verification and explanation verification defines the specific contribution of the paper.

\section{Methodology: Building a System to Automatically Verify the Legality of Public Welfare Distributions}

\subsection{Data Sources}
The Supplemental Nutrition Assistance Program (SNAP) \cite{USDAFNS_SNAP_Overview} is the largest food-assistance program in the United States, providing monthly benefits to low-income households based on statutory eligibility criteria including income, residency, citizenship, resources, and student status. Still, not all eligible individuals receive their benefits \cite{SNAP_stats}. Errors in determination put applicants at risk of losing access to basic subsistence needs.

In California, SNAP operates under the name CalFresh \cite{CDSS_CalFresh}. Eligibility determinations are governed by the California Manual of Policies and Procedures (MPP) Division 63 \cite{mpp_div63_2024}, which translates federal and state requirements into detailed administrative rules. When a decision is made, counties must issue a Notice of Action (NOA) \cite{CDSS_NOA_Docs} explaining the legal basis for an approval, denial, reduction, or termination of benefits. Because internal decision processes may be automated or otherwise opaque, the NOA is often the only publicly visible justification for how eligibility rules were applied. It serves as the central accountability mechanism through which applicants can understand, challenge, or correct decisions affecting their access to food \cite{CDSS_CalFreshRegs}.

Our designed neuro-symbolic framework requires two types of input: (1) A statutory corpus encoding the legal rules that govern eligibility, and (2) Real-world case data containing the agency's explanation, the determination outcome, and the factual circumstances of the applicant. The first provides the legal standard against which explanations are evaluated; the second provides the explanations and case facts to be verified. We draw each from a separate publicly available source. The statutory corpus comes from the CalFresh regulations in MPP Division 63 \cite{mpp_div63_2024}, published by the California Department of Social Services. These span chapters on eligibility criteria, benefit computation, and administrative procedures, and are stored as separate document files on the California Department of Social Services (CDSS) website. We downloaded, read, assembled, and cleaned them into a unified JSON structure for downstream processing. This corpus serves as the ground truth for ontology construction of the law.

Case data was drawn from the CDSS State Hearings Division Decision Registry \cite{cdss_acms_registry}, a public database of administrative appeal decisions. Each record documents a contested CalFresh eligibility determination and includes the agency's original justification, the claimant's grounds for challenge, and the adjudicatory outcome. We downloaded individual hearing decisions as PDFs, cleaned them, and used Python-based extraction scripts with named entity recognition to reconstruct three elements from each case: (1) The explanation quoted from the agency's Notice of Action, (2) The eligibility determination, and (3) Case-related facts presented by the claimant. Because original case files were not directly available, these reconstructed elements serve as the system's input data. We selected 50 cases spanning five statutory eligibility dimensions — income, residency, citizenship, resources, and student status — to ensure coverage of the five primary categories invoked in CalFresh determinations.

\subsection{Overall Architecture}
Our architecture has four stages: Ontology construction, Terminological Box (TBox) encoding, Assertion Box (ABox) construction, and Satisfiability Modulo Theories (SMT)-based verification, as shown in Figure 1. Ontology construction derives a controlled legal vocabulary and statute-based constraints from the governing law of CalFresh. These are assembled into the TBox, which encodes stable, case-invariant legal requirements. In parallel, administrative explanations and case facts are structured into the ABox, capturing what the agency claims as a justification for a particular determination. The verification layer evaluates whether the TBox and ABox are jointly satisfiable using the SMT. A 'SAT' outcome means the explanation is coherent under the governing law i.e. it can function as a legally admissible justification. An 'UNSAT' outcome means no such interpretation exists, signaling that the explanation fails to justify the determination under the law (but not that the eligibility decision itself was wrong).

When an inconsistency is detected, the solver extracts an unsatisfiable core i.e. a minimal subset of constraints responsible for the conflict and maps them back to their corresponding legal provisions. The output is not a binary verdict, but a statute-grounded accountability signal identifying which legal clause is violated by the explanation. Simply put, the ontology and TBox define what the law requires, the ABox captures what the explanation claims, and SMT verification determines whether those claims hold up.

\begin{figure}[H]
  \centering
  \includegraphics[width=0.6\textwidth, height=0.3\textheight, keepaspectratio]{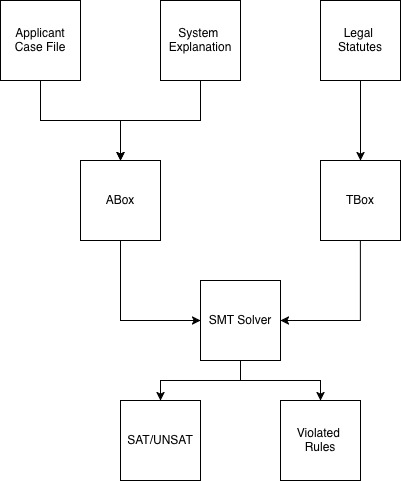}
  \caption{System architecture: case files and explanations are encoded into the ABox, statutory rules into the TBox, and both are evaluated jointly by the SMT solver.}
  \Description{System architecture: case files and explanations are encoded into the ABox, statutory rules into the TBox, and they are jointly evaluated by the SMT solver.}
  \label{fig:architecture}
\end{figure}

\subsection{Ontology Construction}
Statutes and administrative explanations are written in natural language and cannot be checked directly for logical consistency. We therefore construct a formal ontology; a controlled vocabulary of the concepts, relationships, and constraints that govern eligibility decisions to serve as the foundation for verification.

Rather than modeling an entire legal domain, the ontology targets the core elements that structure real-world determinations: threshold conditions, status categories, verification requirements, and exclusion rules. The design prioritizes traceability to authoritative sources and consistency across cases over legal completeness.

The ontology's architecture draws on the LKIF-Core (Legal Knowledge Interchange Format) framework \cite{hoekstra2007lkif}, from which we adopted class naming conventions and structural patterns for representing legal knowledge. The key insight guiding construction is that the structure of the statute itself provides the skeleton of the ontology: the MPP Division 63 regulations are already organized around eligibility dimensions (income, residency, citizenship, resources, student status), and each section specifies the conditions, thresholds, and exceptions that apply within that dimension. We used this statutory structure as the top-level class hierarchy.

Within each statutory section, we used named entity recognition to extract noun-verb keyword pairs. Nouns identified candidate classes and individuals (e.g., "household," "gross income," "sponsor," "verification document"), while verbs identified candidate relationships and properties (e.g., "exceeds," "resides in," "is exempt from"). To prevent redundancy, each candidate term was embedded using the embedding model Qwen3-Embedding-8B \cite{zhang_qwen3_embedding_2025} and compared against existing ontology entries via cosine similarity; terms exceeding a similarity threshold of 0.85 were merged with their existing counterpart rather than added as duplicates. 

\subsection{TBox: Legal Knowledge and Statutory Constraints}
\label{sec:tbox}

The TBox represents the normative structure of the domain. It consists of two parts: a legal ontology that defines the key concepts used in decision-making, and a set of statute-derived rules that specify when a determination is legally permitted or prohibited. Together, these components define the space of legally valid justifications against which individual explanations are evaluated.

The construction of the TBox follows the pipeline shown in Figure 2. The process begins with the regulatory corpus, which serves as the authoritative source of legal requirements. These texts are segmented law by law, preserving the statutory structure of the source material. We then use named entity recognition to extract legally operative noun-verb pairs from each law  identifying conditions, thresholds, obligations, and exclusions. These extracted elements are mapped to the controlled ontology vocabulary using cosine similarity against existing entries, ensuring that the same legal concepts are represented consistently across rules and cases.

Using this shared vocabulary, a large language model, OpenAI's o1 \cite{openai_o1_2024} is prompted to translate each statutory provision into a formal logical rule (see Appendix for prompt strategy). These rules express the normative structure of the law in solver-ready form, including implications, necessary conditions, and disqualifying grounds. Each rule is stored together with a citation to its source provision, preserving a direct link between formal constraints and their legal authority.

This construction process yields a case-invariant TBox: the ontology and rules remain fixed across cases and define the general standards of legal coherence. Individual explanations are later evaluated by instantiating these standards with case-specific facts in the ABox and checking whether the asserted reasoning satisfies the constraints encoded in the TBox.

\begin{figure}[t]
\includegraphics[width=0.6\textwidth, height=0.3\textheight, keepaspectratio]{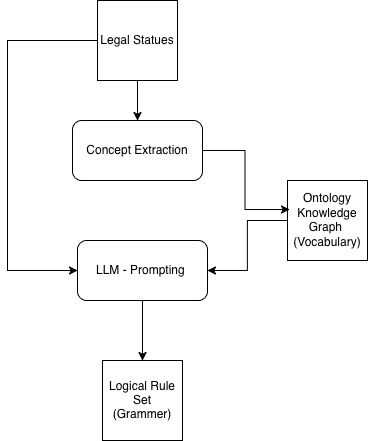}
  \caption{TBox construction pipeline: statutory provisions are segmented, operative terms are extracted and mapped to the ontology vocabulary, and an LLM translates each provision into a solver-ready constraint linked to its source citation.}
  \Description{TBox construction pipeline: statutory provisions are segmented, operative terms are extracted and mapped to the ontology vocabulary, and an LLM translates each provision into a solver-ready constraint linked to its source citation.}
  \label{fig:tbox-pipeline}
\end{figure}

\subsection{ABox: Case Assertions and Explanatory Claims}

\label{sec:abox}
The ABox represents the case-level content of a determination. It encodes two types of information: (i) factual predicates describing the case, and (ii) the claims made by the explanation, including the asserted outcome and the conditions offered in its support. Explanations are mapped into the same controlled predicate vocabulary used by the TBox so that justificatory claims can be evaluated in a consistent and verification-ready form.
The ABox is built using the pipeline shown in Figure 3. As described in Section 3.1, each case is reconstructed from a hearing decision record containing the agency's original explanation, the eligibility determination, and the factual circumstances presented by the claimant. 

The construction of the ABox proceeds in three steps:
First, the justificatory rules cited in the case are fetched from the existing TBox. Because the TBox already encodes the full set of statute-derived constraints, the ABox does not re-derive legal rules. It retrieves the specific provisions that the explanation invokes. This ensures that the same formal representation of the law is used on both sides of the verification.

Second, the agency's explanation is translated into the same statutory logic used by the TBox. OpenAI's o1 \cite{openai_o1_2024} is prompted to encode the explanation's reasoning as solver-ready implication statements (see Appendix for prompt strategy). Although these take the form of logical rules, they do not represent statutory requirements, they capture what the explanation claims would justify the outcome.

Third, the case file is assembled from the hearing decision record itself. Factual predicates such as the applicant's circumstances including income or household size are extracted from the claimant's testimony and case details documented in the hearing decision. These facts, together with the retrieved TBox rules and the translated explanation logic, form a self-contained ABox representing a single case, ready for verification against the TBox constraints.

\begin{figure}[t]
\includegraphics[width=0.6\textwidth, height=0.3\textheight, keepaspectratio]{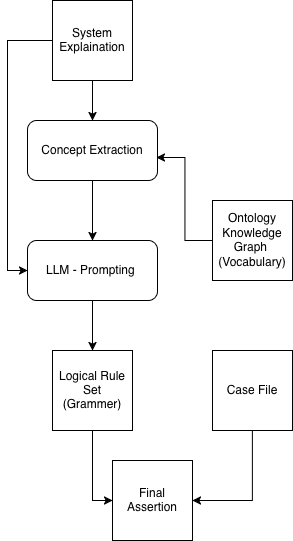}
  \caption{ABox construction pipeline: the agency's explanation is mapped to ontology predicates, translated into solver-ready logic by the LLM, and combined with case facts and retrieved TBox rules to form a complete case assertion.}
  \Description{ABox construction pipeline: the agency's explanation is mapped to ontology predicates, translated into solver-ready logic by the LLM, and combined with case facts and retrieved TBox rules to form a complete case assertion.}
  \label{fig:abox-pipeline}
\end{figure}

\subsection{Solver: Formal Verification and Accountability Output}
\label{sec:solver}
The solver takes the legal rules from the TBox and the case assertions from the ABox and evaluates their joint satisfiability using an SMT solver. Every explanation is treated as a set of testable claims: the asserted determination (eligible or not eligible) and the conditions offered as reasons. Verification asks whether these claims are consistent with the statutory constraints that apply. A SAT result means the explanation and determination are coherent under the law and can function as a legally admissible justification. An UNSAT result means the explanation cannot be made consistent with the legal rules — not that the eligibility decision itself was wrong, but that the justification fails.

To support audit, appeal, and oversight, the solver tracks which legal rules are involved in each check. When an inconsistency is detected, the system extracts an unsatisfiable core i.e. a minimal subset of constraints responsible for the conflict and maps them back to their corresponding provisions in the MPP. The output is not a vague error but a statute-grounded accountability signal, identifying which specific legal requirements the explanation failed to meet. This makes results actionable for review, appeals, and compliance work.

\begin{figure}[htbp]
  \centering
\includegraphics[width=0.6\textwidth, height=0.3\textheight, keepaspectratio]{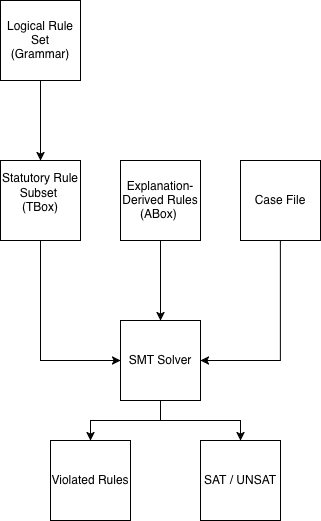}
  \caption{SMT verification pipeline: the TBox rules and ABox assertions are evaluated jointly by the solver, producing a SAT or UNSAT outcome and, when inconsistent, an unsatisfiable core identifying the implicated statutory provisions.}
  \label{fig:smt-pipeline}
  \Description{SMT verification pipeline: the TBox rules and ABox assertions are evaluated jointly by the solver, producing a SAT or UNSAT outcome and, when inconsistent, an unsatisfiable core identifying the implicated statutory provisions.}
\end{figure}

\section{The CalFresh Case Study of Public Algorithmic Governance}
We trace a single case through the full pipeline to illustrate how the system operates (See Figure~\ref{fig:case-walkthrough} in the Appendix). The case involves a one-person household in San Diego County whose CalFresh benefits were terminated (See Figure~\ref{fig:NOA} in the Appendix). The agency's Notice of Action states: ``CalFresh benefits terminated because gross income exceeds the gross income limit for a household of one.''

\paragraph{TBox:} The relevant statutory provision is MPP \S63-409.111, which establishes the gross income eligibility standard: a household with gross income in excess of the standard is ineligible. During TBox construction, this provision was segmented, its operative terms (\texttt{GrossIncome}, \texttt{GrossIncomeLimit}, \texttt{HouseholdSize}) were extracted via NER and mapped to the ontology vocabulary, and o1 translated it into the solver-ready rule:

\begin{center}
\texttt{Implies(GrossIncome > GrossIncomeLimit(HouseholdSize), Not(Applicant\_Eligible))}
\end{center}

\noindent This rule is stored with its citation to \S63-409.111.

\paragraph{ABox:} The case file is assembled from the hearing decision record. Construction follows the three steps described in Section~\ref{sec:abox}. First, the explanation text---``CalFresh benefits terminated because gross income exceeds the gross income limit for a household of one''---is processed using NER to extract its operative noun-verb pairs. This yields the key terms: ``gross income,'' ``exceeds,'' ``gross income limit,'' ``household of one,'' and ``terminated.'' Each term is embedded using Qwen3-Embedding-8B and matched to the ontology vocabulary via cosine similarity, mapping them to the controlled predicates: \texttt{GrossIncome}, \texttt{GrossIncomeLimit}, \texttt{HouseholdSize}, and \texttt{Applicant\_Eligible}.

Second, o1 translates the explanation's reasoning into solver-ready logic using the matched ontology terms:

\begin{center}
\texttt{Implies(GrossIncome > GrossIncomeLimit(HouseholdSize), Not(Applicant\_Eligible))}
\end{center}

\noindent This captures what the explanation \emph{claims} would justify the outcome. The relevant TBox rule citing MPP \S63-409.111 is retrieved---in this case, the explanation's logic and the statutory rule take the same form, indicating close alignment between the agency's stated reasoning and the governing law.

Third, factual predicates are extracted from the hearing decision record: household size of 1, gross earned income of \$728.11, unemployment insurance benefits of \$1{,}950.30, yielding total gross income of \$2{,}678.41 against a gross income limit of \$2{,}082.00. These, together with the translated explanation logic and the retrieved TBox rule, form the complete ABox for this case.

\paragraph{SMT Verification:} The solver evaluates the TBox rule and the ABox assertions jointly. The explanation asserts that gross income exceeds the limit for a household of one, and the determination is termination. The solver returns \textsc{sat}---the explanation is coherent with the cited statutory provision and can function as a legally admissible justification.

To test accountability detection, we invert the determination to ``eligible'' while holding the explanation constant. The solver now returns \textsc{unsat}: an explanation citing excess income cannot justify an approval. The unsatisfiable core identifies MPP \S63-409.111 as the implicated provision---exactly the statute on which the original explanation was grounded. This demonstrates the full accountability loop: the system detects the inconsistency and traces it back to the specific law that was violated.

\subsection {Empirical Examination of the Explanations-based Algorithmic Accountability Framework}
We evaluate the proposed accountability artifact across two layers of the pipeline: 
(1) representation fidelity and (2) Verification and Retrieval of Laws. Together, these analyses assess whether the system can function as accountability infrastructure—carrying legal norms into a computable form, detecting inconsistency between explanations and the governing laws, and localizing statutory responsibility when such inconsistencies arise.

\subsubsection{Representation Adequacy of the Legal Ontology}
We first assess whether the legal ontology provides sufficient structure to support explanation verification. To inspect coverage and internal organization, we embed ontology concepts using a sentence-level embedding model and project them into two dimensions with UMAP \cite{mcinnes_umap_2018} (Figure 5). This visualization is descriptive, it does not affect verification or rule execution. However, it allows us to confirm that the primary statutory dimensions relevant to eligibility form coherent, separable groupings in the representation.

\begin{figure}[htbp]
  \centering
  \includegraphics[width=0.6\textwidth, height=0.3\textheight, keepaspectratio]{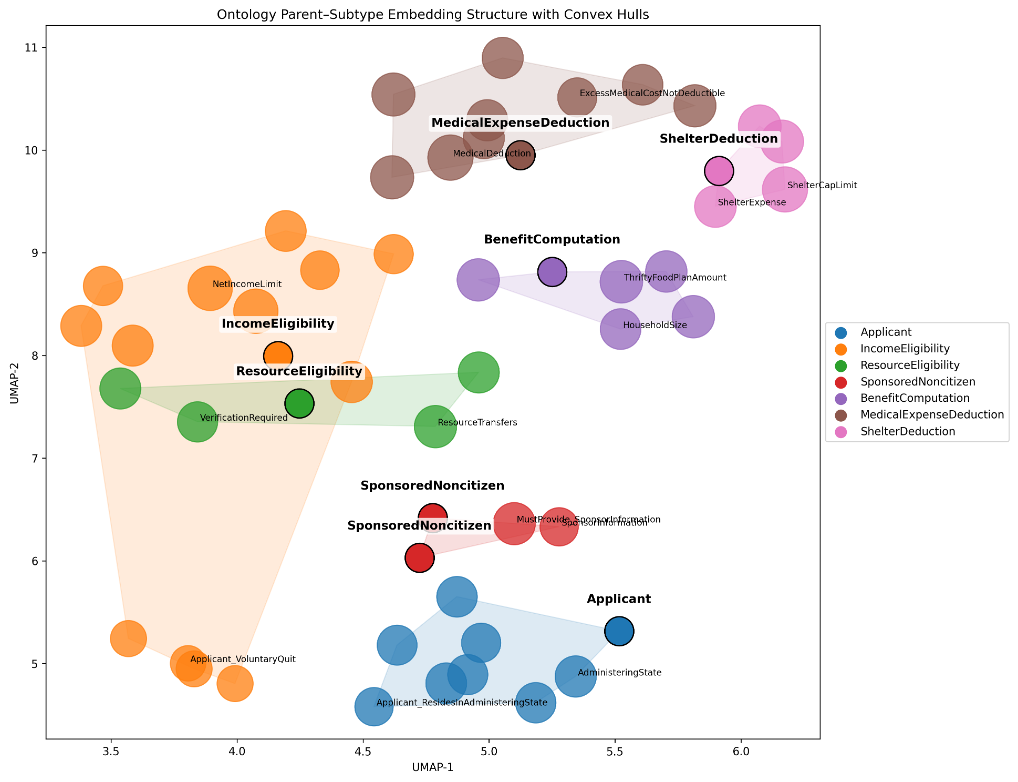}
  \caption{UMAP projection of ontology concepts by eligibility domain. Distinct clusters confirm that legally independent categories remain separable in the representation.}
    \Description{A UMAP scatter plot showing ontology concepts clustered by eligibility domain, with distinct groupings for income eligibility, residency, citizenship, resources, student status, and benefit computation.}
  \label{fig:umap}
\end{figure}

As seen in figure \ref{fig:umap}, across the evaluated CalFresh cases, the ontology captures the primary statutory dimensions invoked in administrative explanations, including income eligibility, residency verification, citizenship status, household composition, resources, and student status. These dimensions align with the categories routinely cited in Notices of Action and eligibility determinations. All cases in our evaluation could be encoded using this vocabulary without resorting to ad hoc predicates, indicating that the representation provides adequate coverage for the accountability task.

This result establishes a necessary precondition for verification: legal requirements can be carried into a formal system in a way that supports consistency checking and statutory traceability. Importantly, this assessment does not claim doctrinal completeness or legal finality. It demonstrates that the ontology is fit for purpose as infrastructure for explanation-level accountability.

\subsubsection{Verification and Retrieval of Laws}
We evaluate whether the accountability artifact can (i) detect when an explanation no longer provides a legally admissible justification for a decision and (ii) localize the statutory provisions implicated when such inconsistency arises. Together, these tests assess the system’s capacity to function as accountability infrastructure for audit, appeal, and oversight.

The evaluation focuses on a controlled subset of cases in which each determination is grounded in a single statutory provision, enabling unambiguous attribution of legal responsibility. For each case, we construct a paired evaluation. In the baseline condition, the explanation is assessed together with its original determination and the relevant statute-derived constraints. In these instances, the solver consistently returns SAT, indicating that the explanation and outcome are jointly coherent under the encoded legal requirements.

To construct negative instances, we reverse the eligibility determination while holding the explanation constant. This produces a deliberate legal contradiction: the same explanation now purports to justify the opposite outcome despite being grounded in a single statutory rule that does not support that determination. This design isolates the accountability function of the artifact by varying only the decision claim rather than introducing speculative factual perturbations.

Under this condition, the solver returns UNSAT whenever the explanation lacks a valid statutory basis for the inverted determination. These UNSAT outcomes do not indicate that the applicant is factually ineligible; rather, they signal that the justification fails as a legal basis for the asserted decision. In institutional terms, the system evaluates not whether the decision is substantively correct, but whether the explanation is legally admissible.

Detection alone, however, is insufficient for accountability. When the solver returns UNSAT, the artifact also localizes legal responsibility by identifying which statutory provisions are implicated. Each statute-derived rule is asserted with a unique identifier corresponding to its source in California's Manual of Policies and Procedures. We extract an unsatisfiable core over these tracked constraints, yielding a subset of statutory provisions that cannot be jointly satisfied with the explanation and asserted determination.

Using the paired single-law cases described above, we evaluate whether the system returns the statute on which the original explanation was based. Because each negative instance is constructed from a case grounded in a single statutory provision, the expected outcome is unambiguous: a legally inconsistent explanation should implicate that same provision when the determination is inverted. Across negative cases, the returned unsatisfiable cores consistently include the expected statute identifier. Moreover, the cores remain compact, typically implicating a small number of provisions rather than producing diffuse attributions. This indicates that the artifact does not merely detect inconsistency but localizes it with sufficient precision to support institutional accountability processes.

To characterize performance, we report two complementary measures: (i) the proportion of negative cases yielding UNSAT outcomes, indicating detection reliability and (ii) the inclusion rate of the expected statute identifier in the returned set of implicated provisions, indicating attribution accuracy. Together, these measures capture the system’s capacity to transform formal verification outcomes into governance-relevant accountability signals.

\subsection{SAT and UNSAT results}
% in preamble:
% \usepackage{multirow}

% in preamble:
% \usepackage{multirow}

\begin{table}[t]
\centering
\small
\begin{tabular}{l c c c c}
\toprule
Category & \# Cases & Status & Localization F1 & SMT Acc. \\
\midrule
Student Status & 10 & SAT   & --   & 0.90 \\
               &    & UNSAT & 0.64 & 0.90 \\
\midrule
Income         & 10 & SAT   & --   & 1.00 \\
               &    & UNSAT & 0.51 & 1.00 \\
\midrule
Residency      & 10 & SAT   & --   & 1.00 \\
               &    & UNSAT & 0.79 & 1.00 \\
\midrule
Citizenship    & 10 & SAT   & --   & 1.00 \\
               &    & UNSAT & 0.83 & 1.00 \\
\midrule
Resources      & 10 & SAT   & --   & 1.00 \\
               &    & UNSAT & 0.72 & 1.00 \\
\midrule
\textbf{Total} & \textbf{50} & -- & -- & \textbf{0.977} \\
\bottomrule
\end{tabular}
\caption{Explanation verification performance across eligibility categories. Each category includes paired coherent (SAT) and incoherent (UNSAT) evaluation instances. Localization F1 is reported for UNSAT cases only.}
\label{tab:verification_by_category}
\end{table}

We evaluate the proposed accountability artifact on a set of 50 administrative cases, constructing a paired incoherent instance for each by inverting the asserted determination, yielding 100 total evaluation instances. Table~1 reports category-level results on a balanced subset of these cases across five statutory dimensions of eligibility. For each category, ten underlying cases are evaluated in both coherent (SAT) and incoherent (UNSAT) form, producing twenty evaluation instances per category.

Across all categories, the system consistently distinguishes legally coherent from incoherent explanations. In coherent cases, verification returns satisfiable outcomes, indicating that the stated justifications align with statutory requirements. In incoherent cases, the solver reliably identifies inconsistencies and retrieves the laws implicated in the explanation. Overall solver behavior remains stable across domains, demonstrating consistent verification under both positive and negative conditions.

Performance in retrieving relevant laws, measured by F1 score on UNSAT cases, varies by category in ways that reflect differences in legal structure. Categories governed by more determinate rules, such as citizenship and residency, exhibit higher retrieval accuracy, while domains characterized by greater contextual nuance, such as income and student status, show lower but still substantial performance. These patterns suggest that the framework is sensitive to the normative complexity of different legal dimensions rather than reflecting uniform system behavior.

Taken together, these results show that the proposed framework supports explanation-level legal accountability along three dimensions: it (i) distinguishes coherent from incoherent justifications, (ii) retrieves the statutory bases of inconsistency when they arise, and (iii) does so consistently across heterogeneous eligibility categories. Rather than optimizing for predictive accuracy, the evaluation demonstrates that the system fulfills its intended institutional role, which is transforming explanations into auditable, legally grounded objects of review. All code, prompts, ontology artifacts, and evaluation data are available at \href{https://github.com/Allen-1242/FAccT_NeuroSymbolicFramework}{\textcolor{blue}{https://github.com/Allen-1242/FAccT\_NeuroSymbolicFramework}}.

\section{Discussion}

The results of this study suggest that explainability in public administration must be reframed as a problem of accountability infrastructure. By treating explanations as normative claims that must be reconciled with statutory requirements, we shift evaluation from model-centric notions of interpretability or bias, to institution-centric notions of governance. This re-framing aligns explanations with their actual role in administrative practice: they are instruments through which legal authority is exercised, contested, and can be corrected. Neuro-symbolic architectures also reflect a broader design principle, in highly regulated domains, automation should prioritize verifiability over autonomy. 

It is important to note that the accountability gap we address does not depend on the presence of machine learning in the eligibility pipeline. Many public benefits systems, including CalFresh, rely on rule-based decision trees or hybrid workflows in which automated screening is combined with caseworker discretion. The opacity of these systems comes from institutional complexity i.e. layered regulations, frequent policy changes, and administrative processes that are difficult for applicants to observe or reconstruct. Whether a determination is produced by a predictive model or a deterministic rule engine, the explanation issued to the applicant must independently satisfy statutory requirements. Our framework evaluates this explanatory layer regardless of the upstream decision mechanism, making it applicable across the full spectrum of automation in public-sector administration.

The designed capability can be used by multiple institutional actors within the eligibility decision life cycle. For agencies, it offers a mechanism for internal compliance checking prior to issuing determinations. For oversight bodies, it provides a structured basis for reviewing contested decisions. For affected individuals and advocates, it opens the possibility of appeals grounded not only in narrative disagreement but in formalized legal inconsistency. In each case, accountability shifts from abstract ethical principles to operational standards tied to the governing laws. Based on evaluation results, agency staff can review their explanations to applicants, public benefits attorneys can seek justice for their clients, and our tool can be routinely used to evaluate the CalFresh program or other programs on a regular basis (rather than upon a specific challenge). The assessments provided by our tool can be monitored by the state comptroller or agency reviewers, who can fine tune the tool upon changes in benefits eligibility criteria.

We envision our framework as a foundational base layer atop which further bands of verification can be composed. For example, privacy-preserving verification modules that formally check whether explanations inadvertently disclose sensitive applicant attributes, or group fairness auditors that assess whether the consistency of reasoning holds equitably across demographic subgroups. By anchoring these higher-order concerns to a verified legal-semantic base, each successive band inherits the formal guarantees established below it rather than operating in isolation.

Ultimately, this work suggests a broader agenda for responsible AI in the public sector. Rather than focusing exclusively on improving model-level explainability, future research should attend to the institutional pathways through which explanations acquire meaning and force. Accountability-centered design asks different questions: not only “can users understand this explanation?” but “can institutions act on it?” and “can it sustain contestation, audit, and review?”

By demonstrating that explanation verification can be operationalized through a neuro-symbolic architecture, this paper offers a concrete step toward that agenda. The contribution is a new kind of infrastructure, one that embeds legal accountability into the technical substrate of public-sector AI. In doing so, it reframes explainability as a governance challenge and positions accountability as a first-class design objective rather than an afterthought.

\section{Limitations}
While the proposed accountability artifact demonstrates the feasibility of explanation-level legal verification, several limitations shape the scope and interpretation of these results. 

\textit{Scope of Legal Formalization}.
First, the system relies on a task-specific formalization of statutory requirements. Welfare law contains open-textured standards, discretionary judgments, and contextual exceptions that cannot be fully captured in symbolic rules or ontological structures. The encoded constraints reflect those aspects of the statute that are routinely invoked in administrative justifications, but they do not constitute a complete or authoritative representation of the law. As a result, the artifact should be understood as supporting accountability for formalized components of legal reasoning, not as a substitute for holistic legal judgment.

\textit{Dependence on Structured Explanations}.
Second, the effectiveness of the system depends on the availability of explanations that can be reliably mapped to the controlled predicate vocabulary. Although neural extraction components mitigate this constraint by enabling scalable processing of unstructured text, errors in extraction or misalignment between explanation language and the ontology can propagate into the verification stage. In practice, this means that the artifact’s outputs are only as robust as the quality of the structured representations it receives.

\textit{Evaluation via Controlled Cases}.
Third, the evaluation emphasizes controlled case constructions—particularly single-law determinations and paired inversions—to enable clear attribution of legal responsibility. While this design is appropriate for testing detection and localization capacity, it does not model the full complexity of real-world administrative decision-making, where multiple statutes, discretionary factors, and procedural considerations often interact. The results therefore demonstrate the system’s capability for accountability verification rather than its performance under all institutional conditions.

\textit{Non-Uniqueness of Legal Localization}.
Fourth, the localization of statutory responsibility is based on unsatisfiable cores returned by the solver. Such cores are not guaranteed to be unique, and different minimal sets of constraints may explain the same inconsistency. Consequently, the artifact should not be interpreted as identifying the singular “true” violated law, but rather as surfacing a principled subset of implicated provisions that warrant institutional attention. This reflects a broader feature of formal verification: it supports accountability by narrowing the space of legal concern, not by resolving interpretive disputes.

\textit{Institutional Integration and Adoption}.
Finally, the artifact has been evaluated as a technical system rather than as part of a deployed administrative workflow. Its real-world impact depends on how it would be integrated into agency practices, oversight regimes, and appeal processes. Questions of organizational uptake, legal admissibility, and procedural legitimacy remain open and require empirical study beyond the scope of this paper.

Taken together, these limitations clarify the intended contribution of the work. The proposed system does not aim to automate legal judgment or guarantee the substantive correctness of administrative decisions. Instead, it offers a proof-of-concept for how neuro-symbolic methods can support accountability infrastructure, making the legal adequacy of explanations visible, reviewable, and contestable within institutional processes.

\section{Conclusion}

Automated decision systems increasingly mediate access to public benefits, rights, and services. In these settings, explanations are not just  technical artifacts but are legal instruments that shape accountability, contestation, and procedural fairness. This paper has argued that evaluating explanations solely in terms of interpretability is insufficient for public-sector AI. What is required instead is an infrastructure that makes explanations legally verifiable.

We introduced a neuro-symbolic accountability artifact that operationalizes this shift. By translating statutory requirements into formal constraints and evaluating administrative explanations against those constraints, the system reframes explainability as a problem of governance rather than transparency alone. Across a multi-layer evaluation we show that explanation-level accountability can be made computationally tractable without automating legal judgment or replacing institutional discretion.

The contribution of this work is a new kind of technical substrate for public administration: one that embeds legal accountability into the design of AI-enabled processes. By making the legal adequacy of explanations visible, inspectable, and contestable, the proposed artifact supports emerging practices of algorithmic auditing, oversight, and rights-preserving AI deployment.

More broadly, this work points toward an accountability-centered paradigm for responsible AI. Rather than treating explainability as an end in itself, future systems should be designed around the institutional roles that explanations are meant to serve—supporting review, appeal, and governance in domains where automation carries legal and social consequences. Neuro-symbolic approaches offer a promising path toward this goal, by strengthening the conditions under which automated systems remain answerable to the law.

\section{Public Repository}
All work including the constructed ontology and the formal generated rule set can be found in \href{https://github.com/Allen-1242/FAccT_NeuroSymbolicFramework}{\textcolor{blue}{\nolinkurl{https://github.com/Allen-1242/FAccT_NeuroSymbolicFramework}}}.

\section{Acknowledgments}
We thank Professor Gabriel Kaptchuk (Department of Computer Science) and Professor Jessica Vitak (College of Information) at the University of Maryland, College Park for their guidance and support.

\section{Generative AI Usage Statement}
The authors used generative AI tools (ChatGPT) during manuscript preparation for limited purposes of language editing, clarity improvements, and structural refinement of text drafted by the authors. No generative AI system was used to produce the substantive content, arguments, analysis, or results of this paper. The authors take full responsibility for the originality, accuracy, and integrity of the work.

\bibliographystyle{ACM-Reference-Format}
\bibliography{references}

\appendix
\section{Appendix}

\begin{figure}[H]
  \centering
\includegraphics[width=1.0\textwidth, height=0.7\textheight, keepaspectratio]{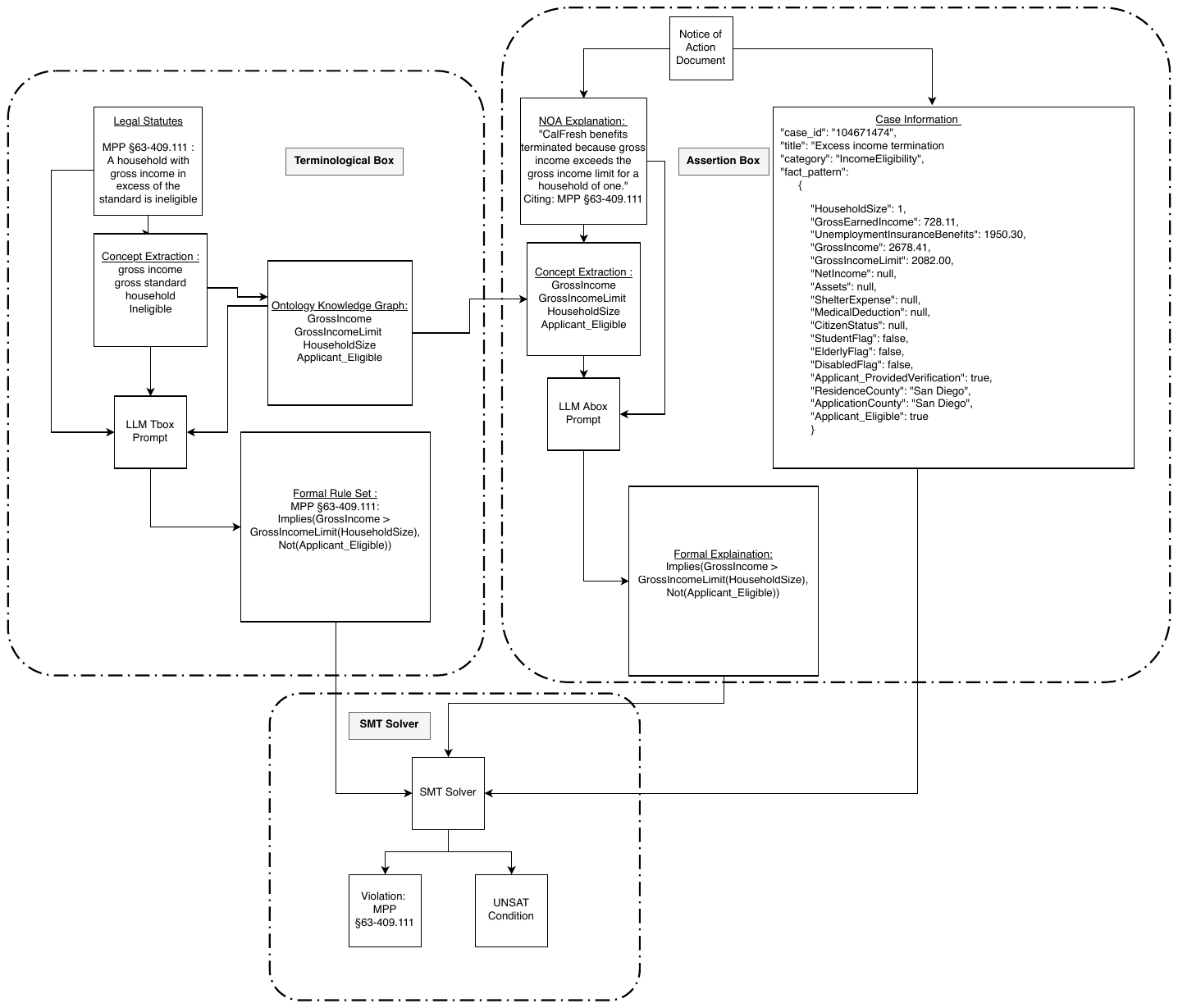}
  \caption{Case Walkthrough with the input provided for the construction of the TBox, ABox, and SMT for this specific benefits determination.}
  \Description{Case Walkthrough with the input provided for the construction of the TBox, ABox, and SMT for this specific benefits determination.}
  \label{fig:case-walkthrough}
\end{figure}

\begin{figure}[H]
\centering
\includegraphics[width=1.0\textwidth, keepaspectratio, page=1]{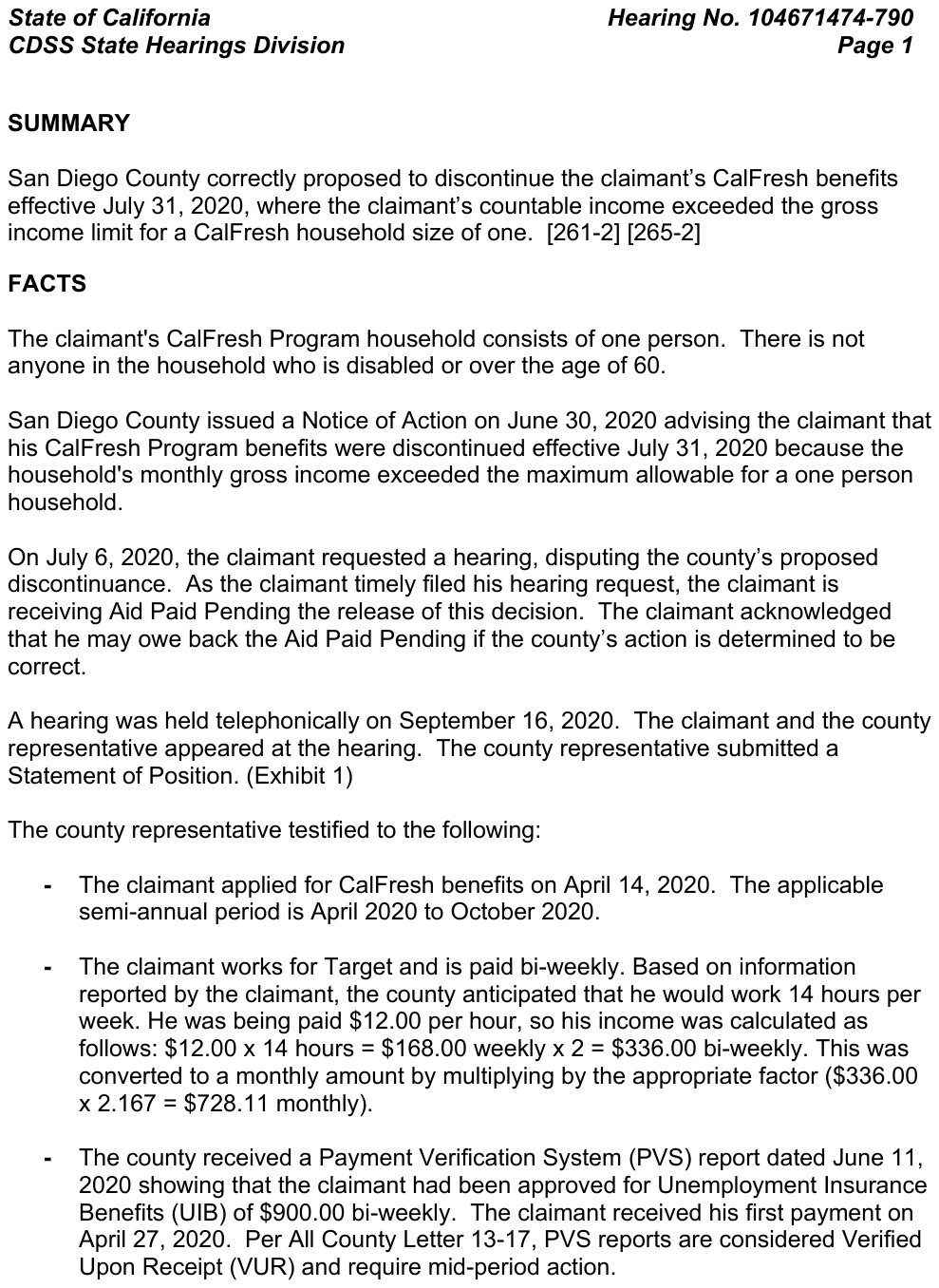}
\caption{Hearing Document (Page 1/3)}
\Description{Hearing Document (Page 1/3)}
\label{fig:NOA}
\end{figure}

\begin{figure}[H]
\centering
\includegraphics[width=1.0\textwidth, keepaspectratio, page=2]{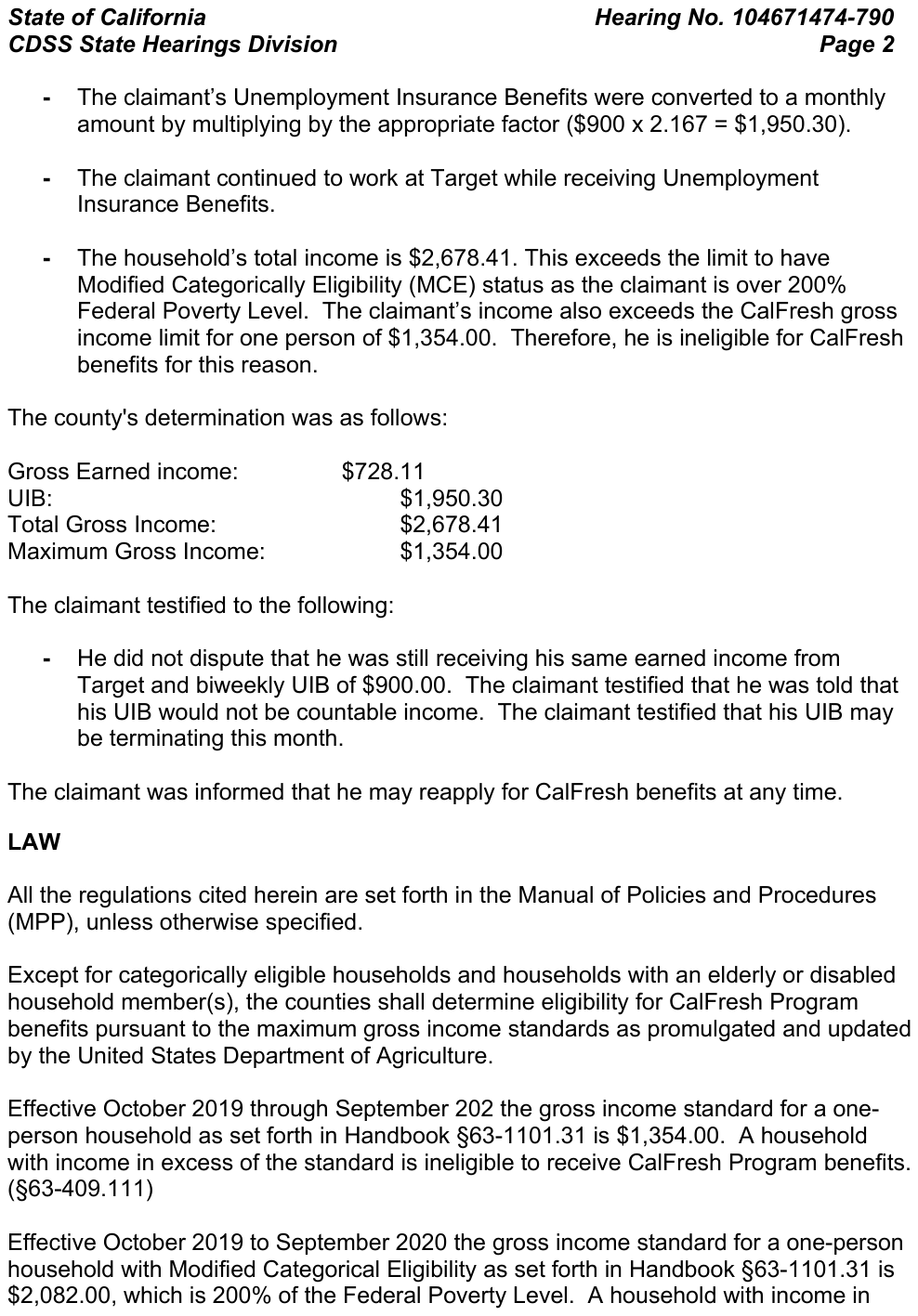}
\caption{Hearing Document (Page 2/3)}
\Description{Hearing Document (Page 2/3)}
\end{figure}

\begin{figure}[H]
\centering
\includegraphics[width=1.0\textwidth, keepaspectratio, page=3]{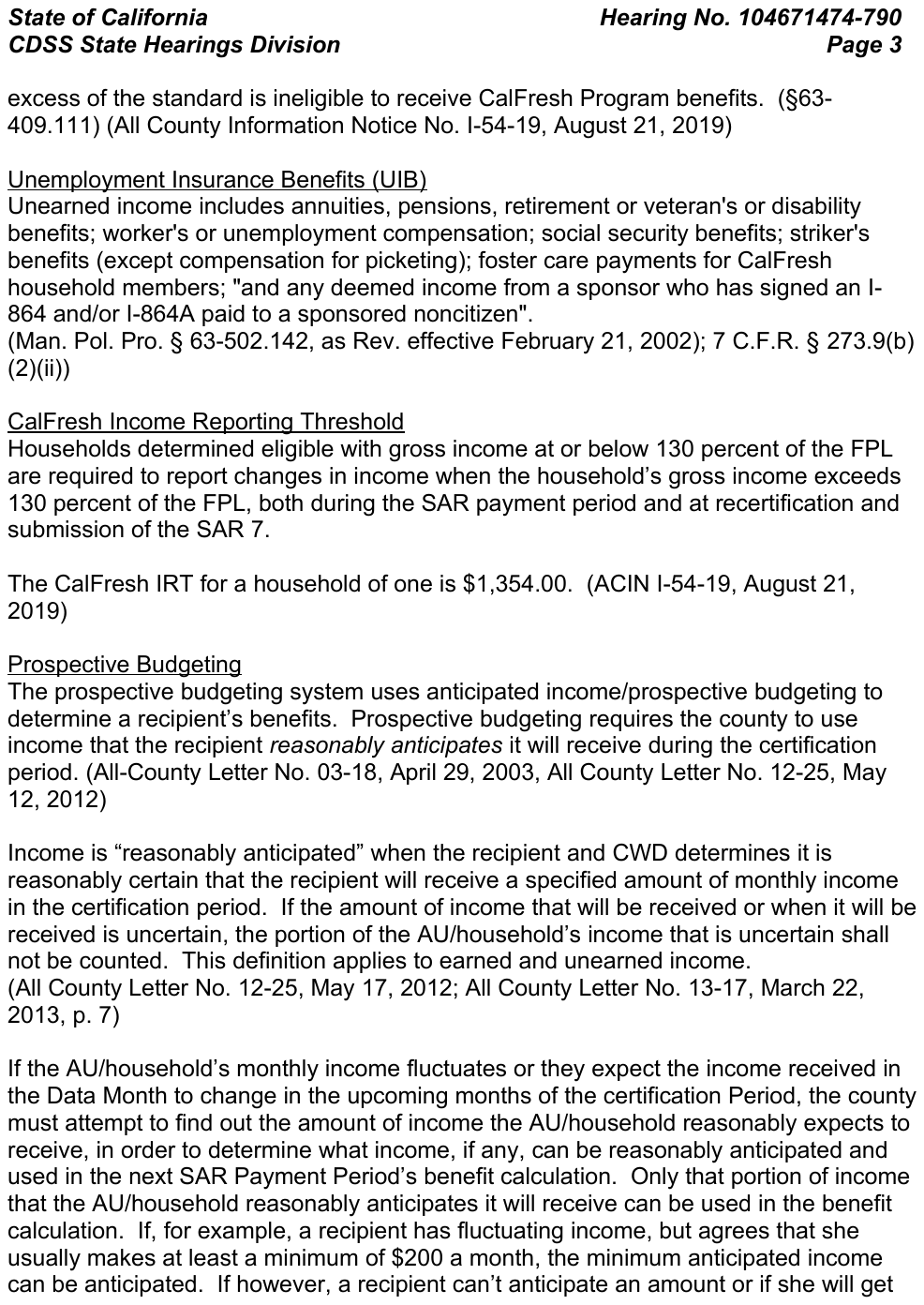}
\caption{Hearing Document (Page 3/3)}
\Description{Hearing Document (Page 3/3)}
\end{figure}

\begin{figure}[h]
\centering
\fbox{\begin{minipage}{0.9\columnwidth}
\ttfamily\small
\noindent You are a legal reasoning assistant that converts
legal clauses into formal logic rules compatible with the
Z3 SMT solver. Use only the ontology variables provided
below. Represent the eligibility conclusion as
Applicant\_Eligible.\\[6pt]
Syntax Requirements:\\
-- Express each rule as a single logical implication\\
-- Use first-order logic operators: Implies, And, Or, Not, Equals\\
-- Use only ontology variable names exactly as listed\\
-- Output only JSON with the field: ``hasLogic''\\
-- Do not include natural language explanations\\[6pt]
Ontology Concepts:\\
\{ontology\_concepts\}\\[6pt]
Clause:\\
``\{input\_clause\}''\\[6pt]
Output format:\\
\{\\
\quad ``hasLogic'': ``<first-order logical implication>''\\
\}
\end{minipage}}
\caption{Prompt template for rule formalization. This template is used for both TBox and ABox construction. For TBox, \texttt{\{input\_clause\}} is a statutory provision from the MPP. For ABox, \texttt{\{input\_clause\}} is the agency's explanation from the Notice of Action.}
\label{fig:prompt}
\Description{Prompt Template for Rule formalization}
\end{figure}

\begin{figure}[H]
\centering
\fbox{\begin{minipage}{0.9\columnwidth}
\ttfamily\small
\noindent\{\\
\quad "IncomeEligibility": \{\\
\quad\quad "definition": "Income factors for eligibility",\\
\quad\quad "conceptType": "Numeric",\\
\quad\quad "citation": "MPP \S63-301",\\
\quad\quad "subtypes": \{\\
\quad\quad\quad "GrossIncome": \{\\
\quad\quad\quad\quad "conceptType": "Numeric",\\
\quad\quad\quad\quad "citation": "MPP \S63-502.11"\\
\quad\quad\quad \},\\
\quad\quad\quad "GrossIncomeLimit": \{\\
\quad\quad\quad\quad "conceptType": "Threshold",\\
\quad\quad\quad\quad "citation": "MPP \S63-301.1"\\
\quad\quad\quad \},\\
\quad\quad\quad "NetIncome": \{\\
\quad\quad\quad\quad "conceptType": "Numeric",\\
\quad\quad\quad\quad "citation": "MPP \S63-301.2"\\
\quad\quad\quad \},\\
\quad\quad\quad "IncomeWithinLimits": \{\\
\quad\quad\quad\quad "conceptType": "Boolean",\\
\quad\quad\quad\quad "citation": "MPP \S63-301.1-2"\\
\quad\quad\quad \}\\
\quad\quad \}\\
\quad \},\\
\quad "Applicant": \{\\
\quad\quad "conceptType": "Entity",\\
\quad\quad "citation": "MPP \S63-401",\\
\quad\quad "subtypes": \{\\
\quad\quad\quad "Applicant\_Eligible": \{\\
\quad\quad\quad\quad "conceptType": "Boolean",\\
\quad\quad\quad\quad "citation": "MPP \S63-401.1"\\
\quad\quad\quad \}\\
\quad\quad \}\\
\quad \}\\
\}
\end{minipage}}
\caption{Excerpt of the ontology in JSON format, showing two of nine domains. Each concept includes a type for solver compatibility and a citation linking it to the governing statute. Definitions omitted for space.}
\Description{Ontology Example in JSON format}
\label{fig:ontology}
\end{figure}

\begin{figure}[H]
\centering
\fbox{\begin{minipage}{0.9\columnwidth}
\ttfamily\small
\noindent[\{\\
\quad "id": "Rule\_GrossIncomeOverLimit",\\
\quad "citation": "MPP \S63-301.1",\\
\quad "hasText": "Households whose gross income exceeds\\
\quad\quad the established limit are ineligible.",\\
\quad "class": "IncomeEligibility",\\
\quad "appliesTo": ["GrossIncome", "GrossIncomeLimit"],\\
\quad "determines": ["Applicant\_Eligible"],\\
\quad "hasLogic": "Implies(GrossIncome >\\
\quad\quad GrossIncomeLimit, Not(Applicant\_Eligible))",\\
\quad "hasModality": "Prohibition"\\
\},\\[4pt]
\{\\
\quad "id": "Rule\_ResidencyRequirement",\\
\quad "citation": "MPP \S63-401.1",\\
\quad "hasText": "A household must reside in the county\\
\quad\quad in which it files an application.",\\
\quad "class": "Residency",\\
\quad "appliesTo": ["Applicant\_ResidenceCounty",\\
\quad\quad "Applicant\_ApplicationCounty"],\\
\quad "determines": ["Applicant\_Eligible"],\\
\quad "hasLogic": "Implies(Applicant\_ResidenceCounty\\
\quad\quad != Applicant\_ApplicationCounty,\\
\quad\quad Not(Applicant\_Eligible))",\\
\quad "hasModality": "Obligation"\\
\},\\[4pt]
\{\\
\quad "id": "Rule\_StudentMustMeetExemption",\\
\quad "citation": "MPP \S63-406.21",\\
\quad "hasText": "Any student shall meet at least one\\
\quad\quad allowable exemption to be eligible.",\\
\quad "class": "StudentEligibility",\\
\quad "appliesTo": ["Applicant\_IsStudent",\\
\quad\quad "Applicant\_HasStudentExemption"],\\
\quad "determines": ["Applicant\_Eligible"],\\
\quad "hasLogic": "Implies(Applicant\_IsStudent == True,\\
\quad\quad Applicant\_HasStudentExemption == True)",\\
\quad "hasModality": "Obligation"\\
\}]
\end{minipage}}
\caption{Excerpt of the TBox rule set showing three rules across income, residency, and student eligibility domains. Each rule preserves its statutory citation, the ontology predicates it operates on, and the solver-ready logic. The full rule set can be examined in the attached github repository.}
\Description{Example of the Tbox}
\label{fig:rules}
\end{figure}

\end{document}